# Wetting layer evolution and its temperature dependence during self assembly of InAs/GaAs quantum dots

Hongyi Zhang, Yonghai Chen,[a] Guanyu Zhou, Chenguang Tang, and Zhanguo Wang

Key Laboratory of Semiconductor Materials Science,

Institute of Semiconductors, Chinese Academy of Sciences,

P.O. Box 912, Beijing 100083, People's Republic of China

**Abstract**

For InAs/GaAs(001) quantum dot (QD) system, the wetting layer (WL) evolution and its temperature dependence were studied using reflectance difference spectroscopy (RDS) and analyzed with a rate equation model. The WL thicknesses showed a monotonic increase at relatively low growth temperatures but a first increase and then decrease at higher temperatures, which were unexpected from the thermodynamic understanding. By adopting a rate equation model, the temperature dependence of QD growth was assigned as the origin of different WL evolutions. A brief discussion on the indium desorption was also given. Those results gave hints of the kinetic aspects of QD self-assembly.

[a] : Electronic mail: yhchen@semi.ac.cn

# 1. Introduction

Epitaxial semiconductor quantum dots (QDs) have attracted much attention because of their application potential in novel optoelectronic devices. [1] They are usually fabricated utilizing the lattice mismatch between the epitaxial layer and substrate, or the Stranski–Krastanov (SK) growth mode. It can be described as follows: For small coverage, two dimensional layer-by-layer growth and the pseudomorphical formation of wetting layer (WL) take place. When the WL reaches a certain critical thickness (CT), a two-dimensional (2D) to three-dimensional (3D) transition starts and the QDs form on the substrate. QDs with high homogeneity in their size and shape are highly advantageous in applications. Basically the WL configuration would also influence the optical properties of QDs and the performance of QD based devices. [2] A controllable growth of QDs with desired properties requires a comprehensive understanding on the growth process. Therefore it is necessary to have a clear understanding of the WL evolution during the QD self assembly.

The commonly accepted thermodynamic understanding of SK mode describes the QD formation on top of a WL of a certain thickness. But it is not accurate for the real situations. It has been reported that in Ge/Si QD system the WL thickness decreases after QD formation. [3-5] It is interpreted in the regime of kinetically controlled QD formation and growth. Since material transfer from WL to QDs sustains the QD formation and growth, a large material consumption rate by QD formation may induce the observed WL erosion. [3,4] As for InAs/GaAs system, a step erosion of WL has also been observed after QD formation. [6] Till now there is no complete description of the WL evolution and its growth condition dependence. In our previous work, reflectance difference spectroscopy (RDS) was used to study the WLs in self assembled nanostructures. Due to its sensitivity, the heavy- and light-hole (HH, LH) related transition energies before and after the QD formation can be directly obtained from the resonant structures in RD spectra. [7-10] In this paper, we studied the WL evolution and its temperature dependence based on the RDS measurements. We found that generally there were two kinds of WL evolution with deposition depending on the growth temperatures. They were well explained in the regime of the temperature dependence of QD growth rate with a rate equation model. The concave up style of evolution was considered as a clear evidence for a non-zero QD growth rate when the WL thickness was smaller than CT. we also gave a simple discussion on the indium desorption during the self assembly. All of these results showed the kinetic aspects of the WL evolution in SK growth.

## 2. Experiments

Six InAs/GaAs(001) QD samples with different growth temperatures (from 490°C to 540°C, with an increment of 10°C) were grown in our Riber-32p MBE system. A gradually changed InAs amount were achieved by stopping the substrate rotation. This method was widely used in studying the QD growth dynamics and to fabricate QD samples with low areal density. [3,5,7,11] The effective indium flux and real deposition amount could be calibrated based on the cosine law for certain configurations of the MBE source beam. [12] Details of the sample growth processes can be found in [10]. To evaluate the WL information, the relative reflectance difference in the sample surface plane, i.e., $\Delta r / r = 2(r_{[110]} - r_{[1\bar{1}0]}) / (r_{[110]} + r_{[1\bar{1}0]})$, was measured with RDS technique in ambient conditions. The setup of our RDS was reported elsewhere. [13]

Figure 1(a) shows the intensity map of the second-derivative RD spectra obtained from the samples grown at 530°C, in which the distinctive features of the GaAs band edge, light hole (LH) and heavy hole (HH) of the WLs can be distinguished. For the horizontal axis the InAs deposition rate is calibrated with the cosine law mentioned above. One can see the LH and HH transition energies redshift almost linearly up to an InAs deposition amount of 1.7ML, which is commonly known as the CT of InAs QD formation, then gradually saturate for further deposition. In order to have an intuitive understanding on the evolution, we calibrate the WL thicknesses based on the transition energies obtained from RD spectra. [14] Figure 1(b) gives the WL thickness evolution of the six samples. The WL thicknesses are signed with open symbols for 2D growth stage and solid symbols for 3D growth stage based on our previous results. [10] In general two distinct evolution processes can be discerned. A first increased and then saturated evolution mode is observed in the samples grown at lower temperatures, while distinct concave up features of the WL evolution for samples with relatively higher growth temperatures appeared right after the QD formation. They are unexpected based on the thermodynamic understanding of the SK growth, for which a stable WL thickness is expected during and after the QD formation. There is another decrease of the WL thickness for the last three samples of each group. The generation of large QDs and dislocation during the QD ripening process, and the enhanced indium absorption abilities by them can be accounted for the decrease. [8,15]

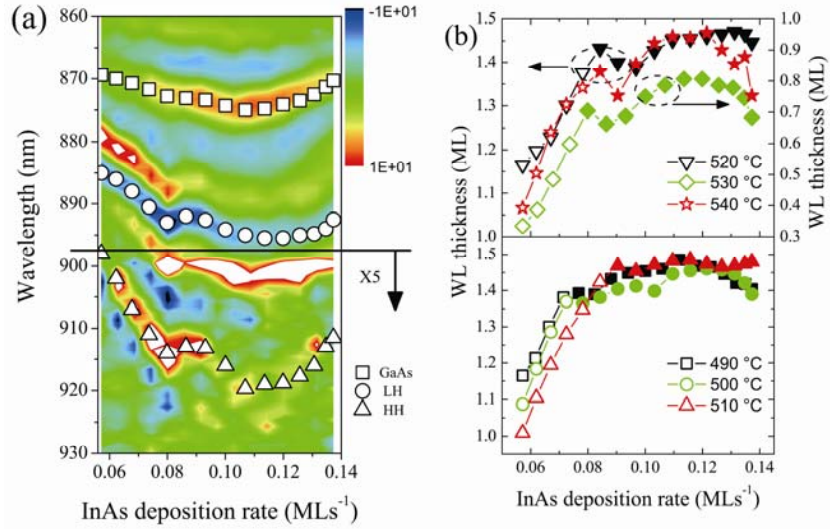

Figure 1 (a) The second-derivative RD spectra ($d^2\rho/d\lambda^2$) of the series of samples grown at 530°C indicated by color contrast. The wavelengths of GaAs band edge, LH- and HH-hole related transitions in the WL are indicated by squares, circles, and triangles, respectively. (b) The WL thickness evolution with InAs deposition amount for samples grown at different temperatures. Note that the WL thicknesses are marked with open symbols at 2D growth stages, and solid symbols at 3D growth stage. The 2D-3D transition points are determined in our previous work. [10]

## 3. Rate equation of the WL thickness

In describing the WL growth dynamics, we consider three main InAs distribution processes. The newly deposited InAs can be incorporated into the WL and QDs, or reevaporated through the indium desorption process. Other processes, such as the formation of quasi-QDs or indium droplets, are neglected for their relatively lower rates. [16,17] In our previous works the CTs of those samples were successfully determined, which enabled us to consider the 2D and 3D growth stages separately in this rate equation model. [10] For 2D growth stage the deposited material contributes to a pseudomorphic growth of the WL and the formation of QDs is neglected. Based on the material balance, the rate equation can be written as,

$$\frac{d\theta}{dt} = G - \frac{\theta}{\tau_{des}}, \quad (t<t_c) \tag{1}$$

where $\theta$ is the WL thickness, $t_c$ is the time of the 2D growth stage, and $G$ is the InAs deposition rate. The indium desorption rate is presumed to be proportional to the InAs amount in WL, and $\tau_{des}$ represents the desorption time constant. The indium desorption process is generally considered as thermal activated.[17,18] $\tau_{des}$ can be written as $\tau_{des} = \frac{1}{v_0}\exp(\frac{E_{des}}{kT})$, where $v_0$ is a preexponential factor and $E_{des}$ is the activation energy of the indium desorption process. By solving equation (1), the WL thickness versus growth time can be written as

$$\theta = G\tau_{des}(1-\exp(-\frac{t}{\tau_{des}})). \quad (t<t_c) \tag{2}$$

Above the CT, a large amount of QDs appear and the newly deposited InAs are mainly consumed by them. The rate equation can be written as

$$\frac{d\theta}{dt} = G - \frac{\theta}{\tau_{des}} - F_{QD}. \quad (t>t_c) \tag{3}$$

Here $F_{QD}$ is used to represent the InAs consumption rate by the QD formation and growth.[19] $F_{QD}$ is determined by the instability of WL and the material diffusion from WL to QDs.[20] The diffusion rate can be written as $D_{In}=(2k_BT/h)\exp(-E_{dif}/k_BT)$,[20] where $k_B$ is Boltzmann's constant, $h$ is Planck's constant, $T$ is the substrate temperature and $E_{dif}$ is the energy barrier. In previous works, for a WL thickness of $\theta$ the instability of WL is commonly considered as $(\theta-\theta_c)$. The driving force of QD growth, which is known as 'superstress', is defined as $\xi=(\theta-\theta_c)/\theta_c$.[21] But it is not suitable in describing our experimental results. The concave up style of evolution shown in the upper panel of figure 1(b) means a non-zero QD growth rate when the WL thickness is slightly below the CT, or else the WL thickness would not reduce below CT in the presence of sufficient InAs supply. A non-stopping QD formation when the WL thickness is smaller than CT is also documented in previous experiments.[22] So here the instable part of WL is written as $(\theta-\alpha\theta_c)$, $(0<\alpha<1)$, correspondingly the 'superstress' is written as $\xi=(\theta-\alpha\theta_c)/\alpha\theta_c$ $(0<\alpha<1)$. The QD formation and growth rate $\gamma$ is considered to be exponentially dependent on the superstress, or $\gamma=b\exp(\beta\xi)$, where $b$ and $\beta$ were constant parameters.[23] Consequently $F_{QD}$ can be written as

$$F_{QD}=b(\theta-\theta_c)\frac{2k_BT}{h}\exp(-E_{dif}/k_BT)\exp(\beta\xi). \tag{4}$$

From equation (3) it is clear that an equilibrium WL thickness is reached when the deposition rate equals to the WL consumption rate by QD formation and indium desorption. The last two processes show strong temperature dependence. So in principle the WL growth have deposition rate and temperature dependences. The WL growth dynamics at different conditions can be obtained from equation (2) and by solving (3) numerically. The calculation results for two different temperatures and varied deposition rates are shown in figure 2. The 2D-3D transition (where the WL thickness exceeds the CT) will not necessarily appear during the growth depending on the deposition rate and time. Generally speaking, firstly the WL thickness shows a nearly linear increase and then is saturated after the QD formation. After that the newly deposited InAs are mainly consumed by the formation of QDs and the WL growth tends to reach equilibrium. The larger the InAs deposition rate is, the thicker the steady-state WL. Concerning the influence of growth temperature, distinct differences can be found by comparing the 3D growth stage in figure 2 (a) and (b). For the low temperature case, the WL thickness increase to equilibrium values which are always above CT. On the other hand, for high temperature grown samples they show WL erosion at the beginning of the 3D growth stage and the WL thicknesses stable at values smaller than CT. The WL erosion disappears by increasing the deposition rates. According to equation (3), the appearance of WL erosion ($d\theta/dt<0$) results from the temperature dependence of QD formation rate, $F_{QD}$. The WL thickness would suffer a decrease if the deposition rate is not large enough to sustain the QD growth at the beginning of 3D evolution stage. Then $F_{QD}$ drops correspondingly according to its dependence on the 'superstress'. It takes some time for the WL thickness to be stable to a certain value, for which the material deposition and the QD growth reach a balance. The bigger the deposition rate is, the thicker the stabilized WL. But if the QD formation rate for the critical WL thickness is lower than the corresponding deposition rate, e.g., a lower growth temperature, the WL thickness would keep on increasing after QD formation, which is the case of figure 2(a).

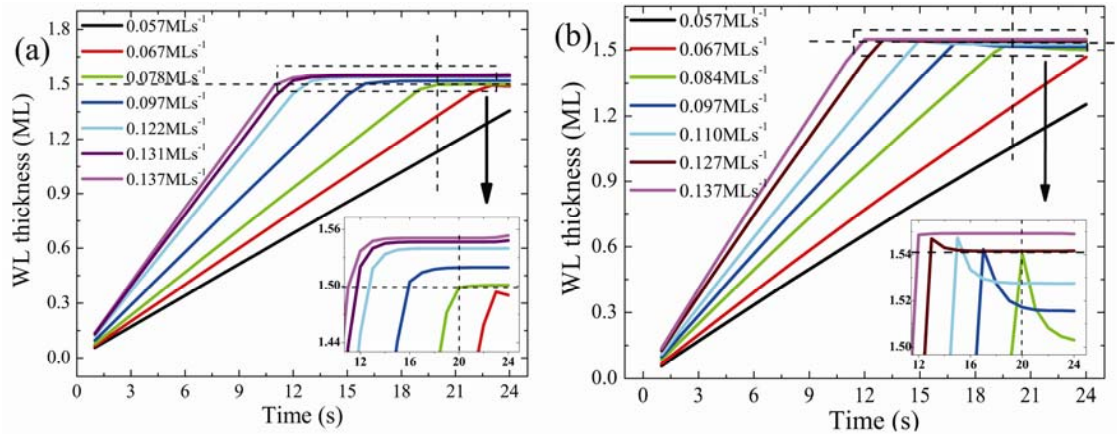

Figure 2: Calculation results of WL growth dynamics for different InAs deposition rates at (a) 490°C and (b) 520°C. The horizontal and vertical dotted lines correspond to the critical thickness and t=20s, respectively. The insets of (a) and (b) zoom in the 3D growth stages.

To understand the experimental results, we have to apply the above mentioned model to the non-rotating samples. The WL thicknesses after deposition can be known by calculating the resulting WL thicknesses with gradually changed InAs deposition rates and a given deposition time. Considering the WL evolution during GI one can simply set $G=0$ in equation (2) and (3) and calculate with the WL thicknesses after deposition as initial values. The simulation results of WL evolution at different temperatures are shown in figure 3(a). One could see that that the main features are well reproduced compared with the experimental results. The WL thickness shows a monotonic increase if the temperature is set at 490°C, but a concave up evolution for the temperatures of 520°C and 540°C. According to the discussion above, we know that the slowed down increase observed at lower temperature is because of the deposition rate dependence of the equilibrium WL thickness. For higher growth temperatures the elevated QD formation rate at the beginning of the 3D growth stage led to the WL erosion, which corresponds to the decrease of WL thickness on those non-rotating samples. The WL thickness increase again with deposition rate when the growth reaches equilibrium. It should also be noted that the simple equations do not reproduce the experimental results quantitatively because of their semi-empirical nature and the use of some adjustable parameters.

We would like to comment on a special feature of those non-rotating samples. The material deposition rate changes gradually at different positions of a sample, which lead to the same behavior of deposition amount for a given growth time. If considering a weak dependence of CT on deposition rate, one would expect that it takes different times at those positions of the sample to enter the 3D growth stage. The 2D growth time $t_{2D}$ can be calculated respectively from equation (2) by taking $\theta=\theta_c$. Then one obtains the 3D growth time $t_{3D}=t_{InAs}-t_{2D}$. The inset of figure 3(a) shows the 3D growth time with deposition rates. It should be noticed that at some positions they are very small values. Apparently a near zero 3D growth time can not ensure an equilibrium quantum dot growth and provide a steady-state WL thickness. It led stronger kinetic-control characters on those samples.

We come back to the 2D growth stage and to study another kinetic problem during growth, the indium desorption. In figure 1(b), for samples with the same deposition amount but different growth temperatures slight differences in the WL thicknesses can be found. From equation (1) we know that it is because of the temperature dependence of indium desorption rate. From equation (2) and further considering the WL evolution during GI, the resulting WL thickness in 2D growth stage can be written as

$$\theta = G\tau_{des}(1-\exp(-\frac{t_{InAs}}{\tau_{des}}))\exp(-\frac{t_{GI}}{\tau_{des}}), \qquad (5)$$

where $t_{InAs}$ is the InAs deposition time and $t_{GI}$ is the GI time. The kinetic parameter of indium desorption, $E_{des}$ and $v_0$, can be extracted from equation (5) and figure 2(b). We adopt the WL thicknesses of the first four groups of samples with effective InAs deposition amounts of 1.14ML, 1.24ML, 1.34ML and 1.45ML to fit $E_{des}$ and $v_0$ respectively. The obtained $E_{des}$ = 3.68eV and $v_0$ are around $5.5*10^{22}$. The activation energy is close to previously reported InAs decomposition energy and indium desorption activation energy from InGaAs. [17,24] We notice that the fitting $v_0$ is such a big number. $v_0$ stands for the attempt frequency of desorption, which is commonly known with the order of $10^{12}$-$10^{14}$ s$^{-1}$ for desorption from metal and semiconductor surfaces. Such a big transition frequency obtained here is also reported by other groups in investigating the InAs/GaAs QD desorption [24], or As desorption from GaAs surface [25], It is considered as physically achievable and could explain several characteristic features in InAs MBE growth. [25] The inset of figure 4 shows the temperature dependence of the desorption life time ($\tau_{des}$) for samples with different InAs deposition amount based on the fitting results. The time constants show a weak dependence on the indium flux, but strongly decrease with increasing temperature. $\tau_{des}$ decreases from 1063s at 490°C to 35s at 540°C for samples with deposition amount of 1.45ML. The same strong dependence is also mentioned elsewhere [26]. Those time constants could be used to estimate the degree of desorption during the growth of InAs/GaAs(001) QDs at a certain temperature.

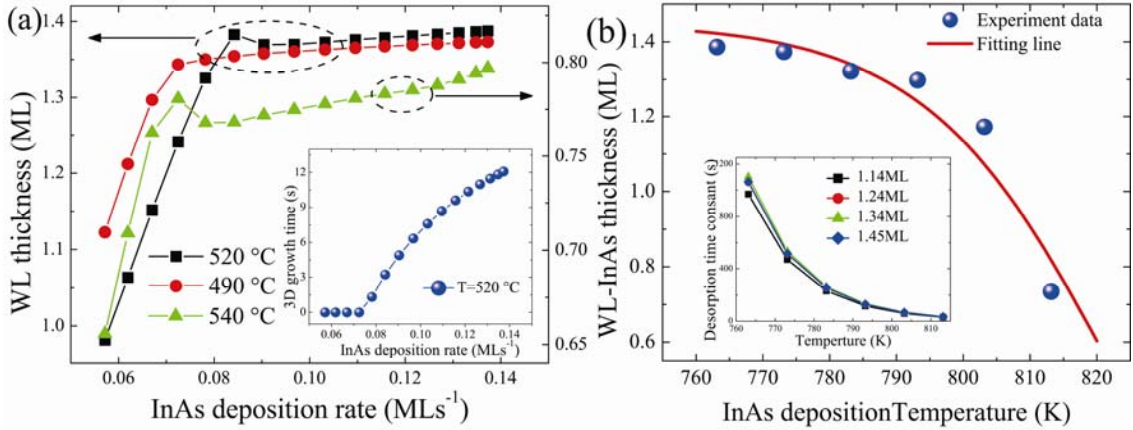

Figure 3: (a) Simulation of the WL evolution of the non-rotating samples at three different temperatures. The open and closed symbols stand for 2D and 3D growth stages respectively. Here we use $b=4\times10^{-9}$, $\alpha=0.8$, $\beta=10$ and $E_{dif}=1.04eV$ to solve the rate equations numerically. The inset of (a) is the dependence of 3D growth time on InAs deposition rate for a given deposition time of 20s. (b) The fitting result of the temperature dependence of WL thicknesses for the sample with a nominal InAs deposition amount of 1.45ML. The inset shows the temperature dependence of the desorption time constants for samples with different InAs deposition rates.

## 4. Conclusion

In conclusion, two kinds of WL evolution process of InAs/GaAs(001) QD system have been discussed based on RDS measurements and a rate equation model. They were well understood in the regime of material balance of WL growth/consumption and the temperature dependence of QD formation. The concave up style of evolution is also an evidence of a non-zero QD growth rate when the WL thickness was slightly lower than the critical value. We also gave a brief discussion on the indium desorption process during growth. Those results helped us in understanding the kinetically controlled the QD growth process.

**Acknowledgement**

The work was supported by the National Natural Science Foundation of China (No. 60990313), the 973 program (2012CB921304, 2012CB619306) and the 863 program (2011AA 03A 101).

**Reference**

1. D. Bimberg, M. Grundmann, and N. N. Ledentsov, *Quantum Dot Heterostructures*. (Wiley, Chichester, 1999).
2. S. Sanguinetti, M. Henini, M. Grassi Alessi, M. Capizzi, P. Frigeri, and S. Franchi, Physical Review B **60** (11), 8276 (1999); L. J. Wang, V. Krapek, F. Ding, F. Horton, A. Schliwa, D. Bimberg, A. Rastelli, and O. G. Schmidt, Physical Review B **80** (8), 085309 (2009); D. G. Deppe and D. L. Huffaker, Applied Physics Letters **77** (21), 3325 (2000); D. R. Matthews, H. D. Summers, P. M. Smowton, and M. Hopkinson, Applied Physics Letters **81** (26), 4904 (2002).
3. R. Bergamaschini, M. Brehm, M. Grydlik, T. Fromherz, G. Bauer, and F. Montalenti, Nanotechnology **22** (28), 285704 (2011).
4. A. V. Osipov, F. Schmitt, S. A. Kukushkin, and P. Hess, Appl. Surf. Sci. **188** (1–2), 156 (2002).
5. Moritz Brehm, Francesco Montalenti, Martyna Grydlik, Guglielmo Vastola, Herbert Lichtenberger, Nina Hrauda, Matthew J. Beck, Thomas Fromherz, Friedrich Schäffler, Leo Miglio, and Günther Bauer, Physical Review B **80** (20), 205321 (2009).
6. E. Placidi, F. Arciprete, V. Sessi, M. Fanfoni, F. Patella, and A. Balzarotti, Applied Physics Letters **86** (24), 241913 (2005); F. Arciprete, E. Placidi, V. Sessi, M. Fanfoni, F. Patella, and A. Balzarotti, Applied Physics Letters **89** (4), 041904 (2006).
7. Y. H. Chen, P. Jin, L. Y. Liang, X. L. Ye, Z. G. Wang, and Arturo I. Martinez, Nanotechnology **17** (9), 2207 (2006).
8. Y. H. Chen, J. Sun, P. Jin, Z. G. Wang, and Z. Yang, Applied Physics Letters **88** (7), 071903 (2006).
9. G. Y. Zhou, Y. H. Chen, J. L. Yu, X. L. Zhou, X. L. Ye, P. Jin, and Z. G. Wang, Applied Physics

Letters **98** (7), 071914 (2011).
10  G. Y. Zhou, Y. H. Chen, C. G. Tang, L. Y. Liang, P. Jin, and Z. G. Wang, J. Appl. Phys. **108** (8), 083513 (2010).
11  Jin Peng, X. L. Ye, and Z. G. Wang, Nanotechnology **16** (12), 2775 (2005).
12  M. A. Herman and H. Sitter, *Molecular Beam Epitaxy: Fundamental and Current Status*. (Springer, Berlin, 1989).
13  Y. H. Chen, X. L. Ye, J. Z. Wang, Z. G. Wang, and Z. Yang, Physical Review B **66** (19), 195321 (2002).
14  M. Geddo, M. Capizzi, A. Patane, and F. Martelli, Journal of Applied Physics **84** (6), 3374 (1998).
15  I. Daruka and A. L. Barabasi, Phys. Rev. Lett. **79** (19), 3708 (1997).
16  R. Heitz, T. R. Ramachandran, A. Kalburge, Q. Xie, I. Mukhametzhanov, P. Chen, and A. Madhukar, Phys. Rev. Lett. **78** (21), 4071 (1997).
17  T. Mozume and I. Ohbu, Japanese Journal of Applied Physics **31** (10), 3277 (1992).
18  C. Heyn, D. Endler, K. Zhang, and W. Hansen, Journal of Crystal Growth **210** (4), 421 (2000).
19  A. Osipov, S. Kukushkin, F. Schmitt, and P. Hess, Physical Review B **64** (20), 205421 (2001).
20  Harvey T. Dobbs, Dimitri D. Vvedensky, Andrew Zangwill, Jonas Johansson, Niclas Carlsson, and Werner Seifert, Physical Review Letters **79** (5), 897 (1997).
21  V. G. Dubrovskii, G. E. Cirlin, and V. M. Ustinov, Physical Review B **68** (7), 075409 (2003).
22  C. Kim, Journal of Crystal Growth **214**, 761 (2000); H. Song, T. Usuki, Y. Nakata, N. Yokoyama, H. Sasakura, and S. Muto, Physical Review B **73** (11), 115327 (2006); A. Tonkikh, G. Cirlin, V. Dubrovskii, Yu Samsonenko, N. Polyakov, V. Egorov, A. Gladyshev, N. Kryzhanovskaya, and V. Ustinov, Technical Physics Letters **29** (8), 691 (2003).
23  A. Balzarotti, Nanotechnology **19** (50), 505701 (2008); A. A. Tonkikh, V. G. Dubrovskii, G. E. Cirlin, V. A. Egorov, V. M. Ustinov, and P. Werner, Physica Status Solidi B-Basic Research **236** (1), R1 (2003).
24  C. Heyn, Physical Review B **66** (7), 075307 (2002).
25  C. Sasaoka, Y. Kato, and A. Usui, Applied Physics Letters **62** (19), 2338 (1993).
26  Ch Heyn and W. Hansen, Journal of Crystal Growth **251** (1-4), 218 (2003).

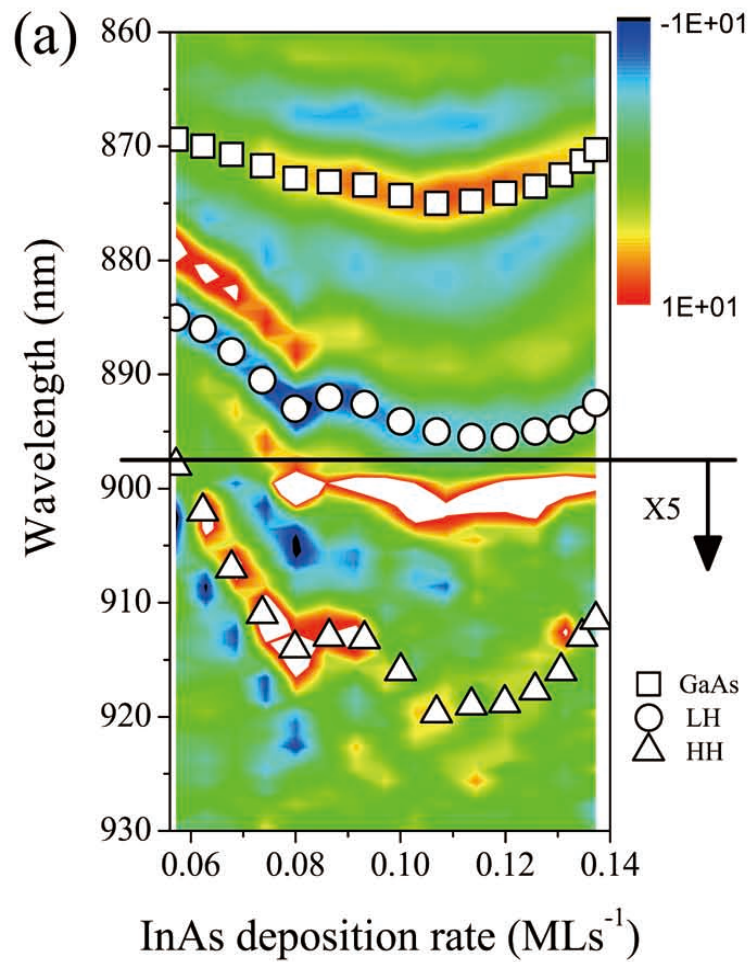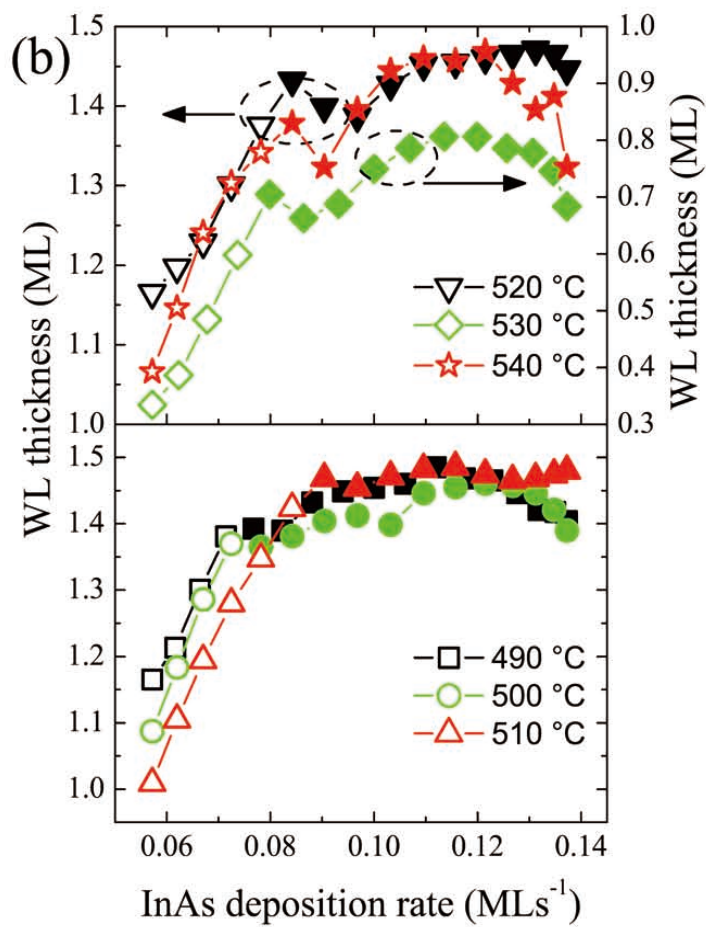

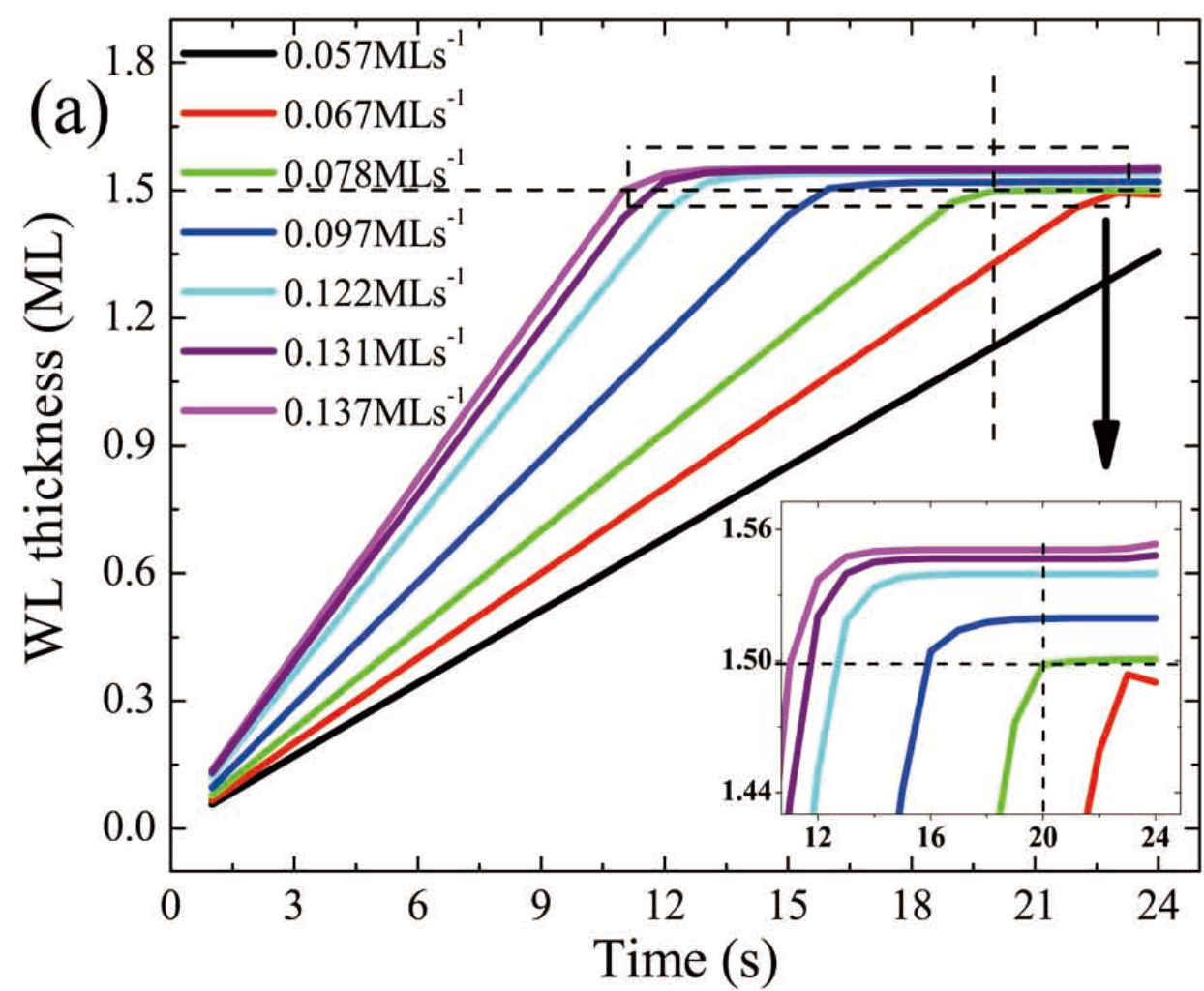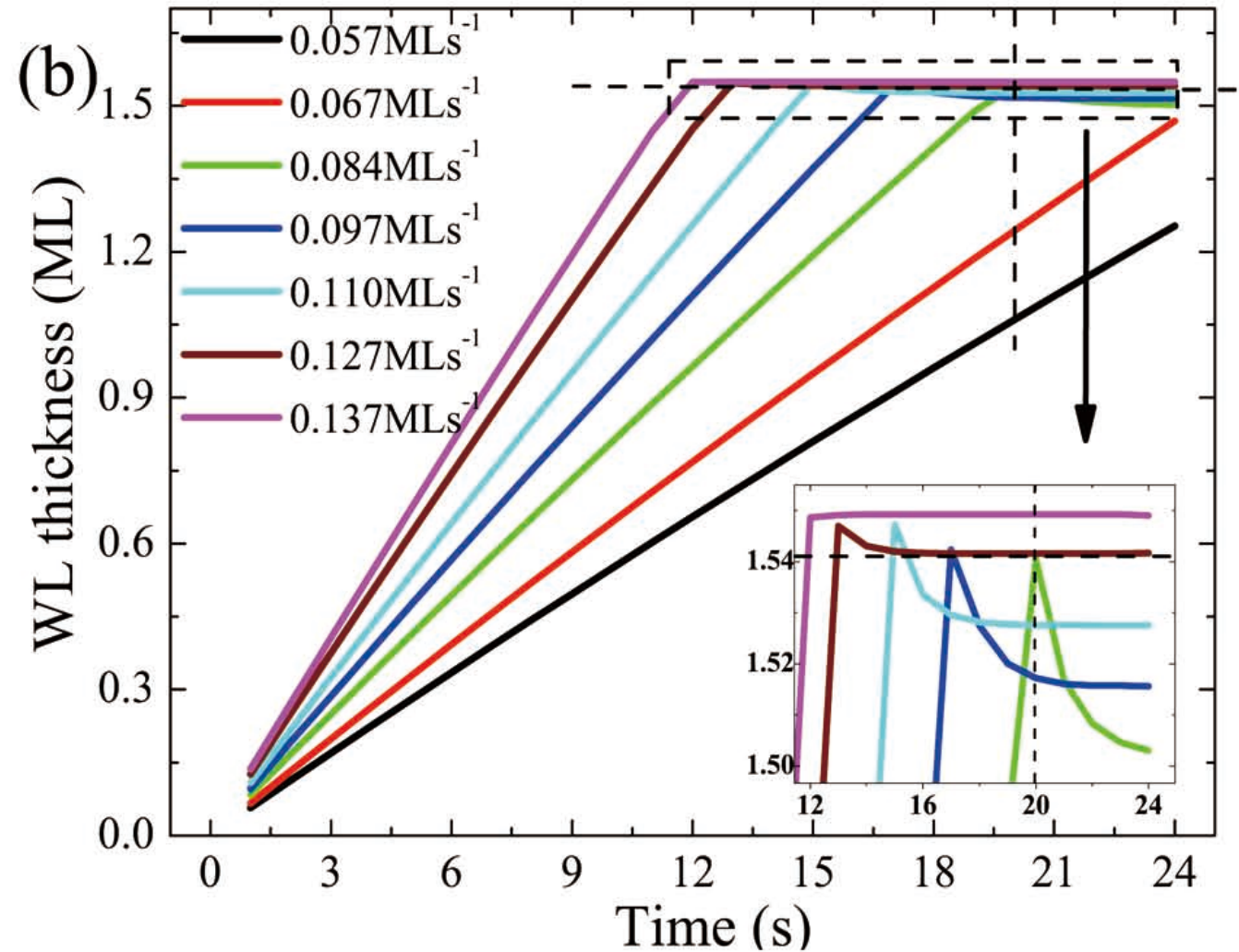

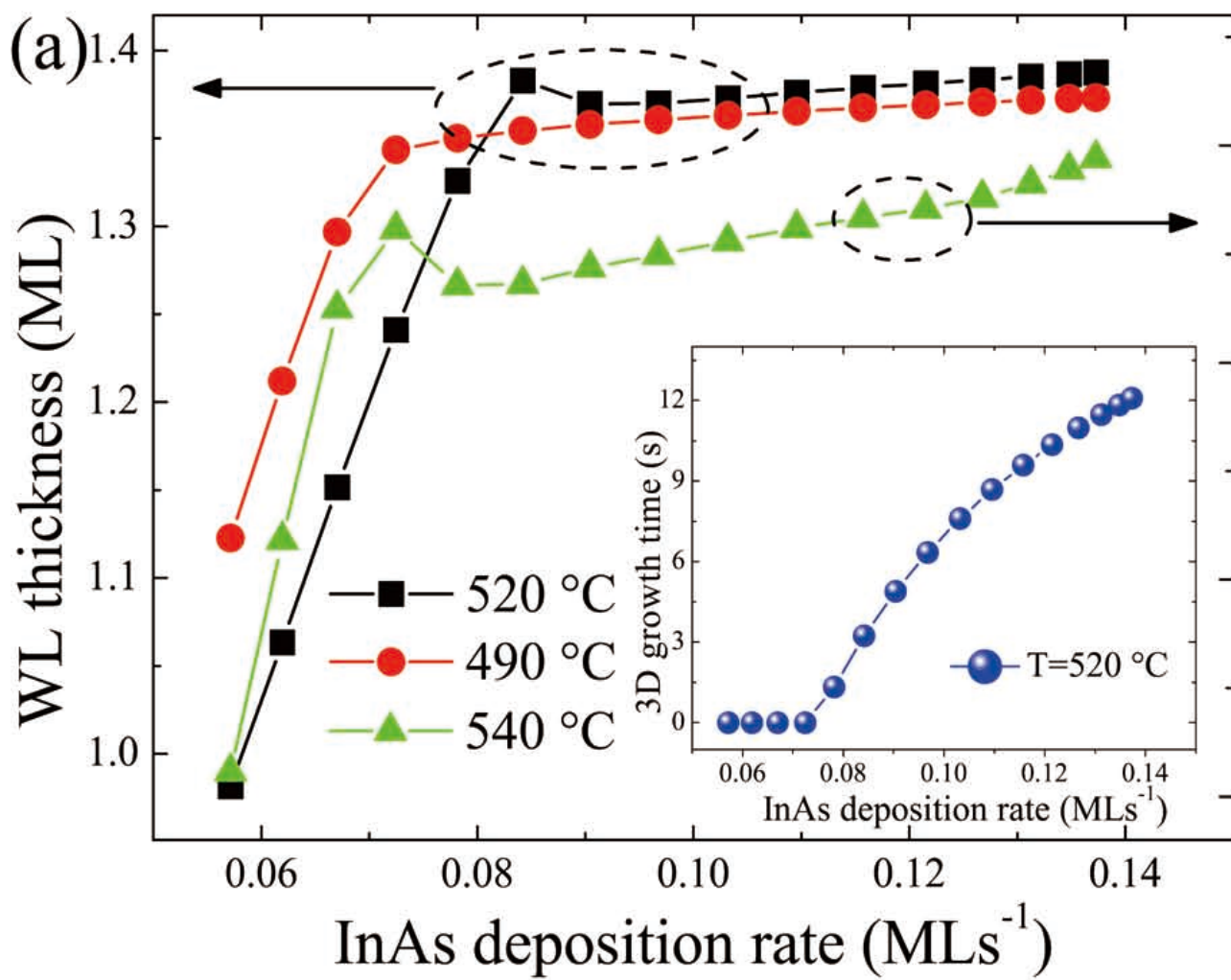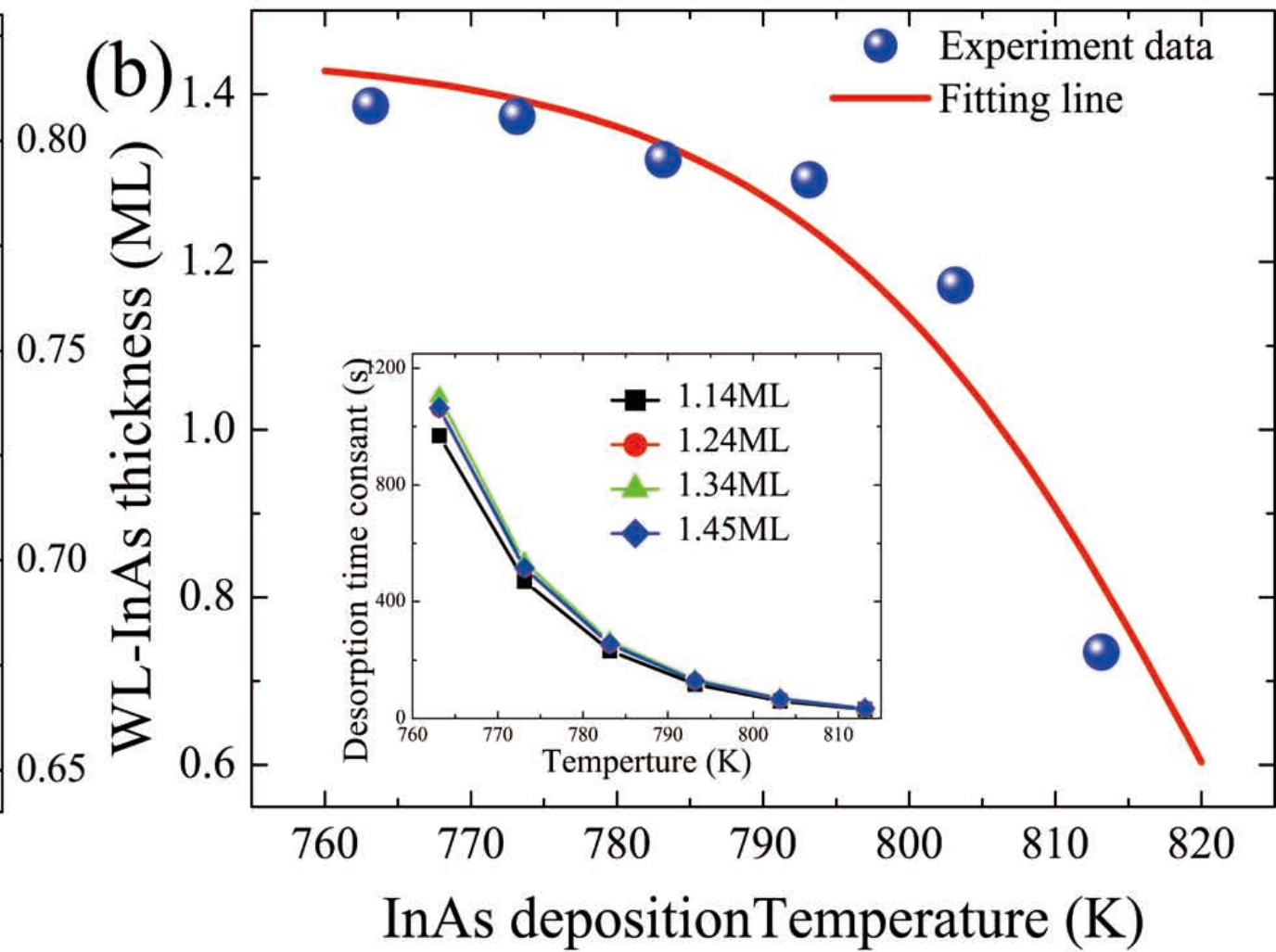